# Concurrent Multiphysics and Multiscale Topology Optimization for Lightweight Laser-Driven Porous Actuator Systems

Musaddiq Al Ali[1] and Masatoshi Shimoda[1]

[1]Department of Advanced Science and Technology, Toyota Technological Institute, 2-12-1

Hisakata, Tenpaku-ku, Nagoya, Aichi 468-8511, Japan

\* Corresponding author: alali@toyota-ti.ac.jp (Musaddiq Al Ali)

**Abstract**

In this research, multi-physics topology optimization is employed to achieve the detailed design of a lightweight porous linear actuation mechanism that harnesses energy through laser activation. A multiscale topology optimization methodology is introduced for micro- and macroscale design, considering energy dissipation via heat convection and radiation. This investigation meticulously considers the impact of heat dissipation mechanisms, including thermal conduction, convection, and radiation. Through various numerical cases, we systematically explore the influence of micro-scale considerations on porous design and understand the effects on the topology optimization process by incorporating various microstructural systems. The results demonstrate that porous actuator designs exhibit superior performance compared to solid actuator designs. This study contributes to advancing the understanding of multiscale effects in topology optimization, paving the way for more efficient and lightweight designs in the field of laser-activated porous actuators.

**1. Introduction**

The advancement in the lithography process, particularly in the progress of 5-nanometer feature processes, has driven the integration of linear electromagnetics as precision actuators [1]. This integration is motivated by the increasing demand for high-precision actuators, driven by the rapid growth of high-end applications such as the production of high-density processing units and nano-manufacturing for applications like quantum computing [2]. Additionally, emerging fields like atomic 3D printing utilizing technologies such as atomic layer deposition [3] have emphasized the requirements for extremely high-precision actuators [4].

In this context, compliance mechanisms, a distinct class of designed mechanical systems with controlled flexibility, offer unique solutions [5]. These compliance mechanisms excel in applications where the integration of conventional rigid mechanisms is difficult or impractical. One notable advantage of compliance mechanisms is their ability to adapt appropriately to changes in specified load conditions. Taking these considerations into account, our ongoing research is dedicated to exploring innovative actuator designs, with a particular focus on laser-activated porous actuators, to address the requirements of high-precision and compact size for advanced technological applications.

The integration of Micro-Electro-Mechanical Systems (MEMS) is a crucial development in the rapidly evolving landscape of advanced technology. MEMS technology adjusts the synergistic effects of miniaturized electronic and mechanical components, generating unparalleled potential across various domains. This presentation meticulously examines the transformative impact specific to MEMS, navigating the complex coexistence of precision and adaptability in describing its multifaceted applications, technological advancements, and the wonders of these small-scale technologies. As we delve into the profound aspects of MEMS technology, let us collectively enjoy this intellectual adventure where precision intertwines seamlessly with miniaturization, redefining the epistemological boundaries of modern engineering and innovation.

For applications in nanofabrication and microsurgery, incorporating compliant actuation mechanisms into the realm of Micro-Electro-Mechanical Systems (MEMS) raises critical considerations. It is essential to miniaturize these mechanisms for seamless remote energy transmission and eliminate the need for a direct power source [6]. The process of miniaturization plays an extremely important role in achieving this specific purpose. Not only does it enhance the efficiency of energy transfer, but it also aligns with the inherently compact nature of MEMS while conforming to the essential goal of achieving improved efficiency in laser-activated compliant actuation mechanisms. Over considerable distances. In many scenarios, especially those in enclosed or inaccessible environments, direct power supply becomes impractical, if not entirely impossible. Therefore, the mission becomes the meticulous design of smaller and streamlined mechanisms. By reducing the physical dimensions of these mechanisms, energy consumption is correspondingly reduced, enabling the utilization of long-lasting energy sources with low power. Such alternative power sources may include batteries and innovative environmental power generation technologies.

This investigation places particular emphasis on the application of laser technology as an extremely accurate optical-thermal energy source [7][8]. Our scrutiny delves into a distinctive subset of compliance mechanisms, specifically examining mechanisms responsive to laser activation. The complexity of utilizing lasers as an energy source for activation within compliant mechanisms constitutes a key focus of our research, elucidating subtle interactions and dynamic responses at the intersection of advanced materials and laser-induced thermal effects.

These mechanisms typically feature spatial configurations that allow changes in the layouts or dimensions of materials when exposed to laser activation. Layout changes are carefully designed to induce specific movements or displacements. Laser-activated compliance mechanisms offer several advantages compared to traditional compliance systems. They exhibit high responsiveness to changes in laser and can be triggered rapidly and easily. Furthermore, they can be designed to achieve advanced precision and reproducibility. Laser-activated compliant mechanisms find applications in various industries.

Despite the myriad benefits and merits intrinsic to compliant mechanisms, their integration into ultra-precision contexts poses a formidable challenge. One primary impediment concerns the attainment of precise motion control alongside the mitigation of undesired deformations that could detrimentally affect overall efficacy. Another formidable barrier, particularly evident when employing laser technology as the activating agent, revolves around the accurate modeling and prediction of compliant mechanism operations. Consequently, the task of designing and fine-tuning these mechanisms for specific applications becomes a formidable endeavor. Moreover, striking an optimal equilibrium between minimizing the mass of compliant mechanisms while upholding their operational efficacy presents a formidable challenge, particularly in pioneering domains such as microelectronics. Additionally, achieving high-fidelity displacement poses challenges without comprehensive consideration of environmental influences [9].

Tackling these obstacles has catalyzed continual research endeavors aimed at investigating novel design methodologies and harnessing sophisticated modeling and simulation techniques to augment the efficiency and dependability of compliant mechanisms in high-precision contexts. Within this realm of innovative design methodologies, topology optimization emerges as a prominent solution. Topology optimization is a robust strategy for creating compliant mechanisms, characterized by its ability to systematically and efficiently navigate vast design possibilities. In this process, numerical methods are used to fine-tune the arrangement and shape of the mechanism, ensuring it meets predefined performance benchmarks while conserving material consumption. This groundbreaking methodology yields compliant mechanisms that exhibit a remarkable balance between lightweight construction and exceptional efficiency [10].

Topology optimization, originating from Michell's foundational work [11] and stemming from J. Maxwell's pioneering research in 1869 [12], has progressed alongside advancements in numerical methods and computational capabilities. Early applications of topology optimization to compliant mechanism design, as demonstrated by Ananthasuresh et al., were primarily based on homogenization principles [13]. However, these initial endeavors often resulted in compliant mechanisms with insufficient flexibility, addressing average design rather than practical applicability. Sigmund introduced an alternative approach, modeling output load as a spring and emphasizing the workpiece's behavior at the output port [14], while Frecker et al. proposed a technique maximizing the ratio of mutual energies, offering a distinct perspective on compliant mechanism design [15][16]. Saxena and Ananthasuresh adopted path-generating mechanisms as guiding principles [17], and Nishiwaki et al. introduced a homogenization-based optimization tailored for enhanced flexibility [18]. These methodologies typically employ multi-objective functions to quantify flexibility, demonstrated through numerous numerical examples [19].

Various methodologies have demonstrated efficacy in compliant mechanism design, particularly under directly applied forces. These include microstructure-based homogenization [20], SIMP interpolation [21], level set ensemble [22], and the ESO method. Zuo et al. introduced an efficient multi-material topology optimization technique, the Ordered SIMP Multi-Material Interpolation Method [23], which enhances computational efficiency by eliminating the need for additional variables. Chu et al. proposed a stress-constrained optimization approach based on level sets, enabling simultaneous control over displacement, compliance, and stress [24]. Rostami et al. analyzed optimal topologies for multi-material compliant mechanisms using the regularized projected gradient approach [25]. Sivapuram et al. presented a multiscale optimization method enabling simultaneous optimization of structure and material at macro and micro scales [26]. Research into thermal expansion as a means of activating compliant mechanisms has advanced significantly. Jonsmann et al.'s foundational work elucidates the thermo-mechanical interactions underlying thermal expansion, informing the development of precisely designed compliant mechanisms [27]. Yin and Ananthasuresh proposed systematic methodologies for topological optimization of electrothermally actuated compliant mechanisms, highlighting the importance of convection modeling [28]. Ansola et al. investigated topology optimization strategies for thermally actuated compliant mechanisms, employing an additive element strategy [29].

As mentioned earlier, the process begins with the absorption of energy from a laser, which then transforms into thermal energy in metals. The dissipation of energy through thermal radiation is a significant physical phenomenon that profoundly influences the design of topologically optimized structures for various applications. However, there has been a relative shortage of research addressing the challenges involving the integration of radiative boundary conditions. This represents an underexplored area in problem-solving.

Castro et al. [30] delved into the design of radiative enclosures using a solid isotropic material with penalization (SIMP) approach to redistribute reflective material. In this framework, 2D element density variables were exploited to interpolate the surface reflectivity inside the enclosure, allowing the determination of temperature and heat flux fields. Munk et al. [31] applied an evolutionary structural topology optimization method to a three-dimensional model of a hypersonic wing, considering aero-thermo-elastic effects. Topology optimization formulations were proposed for the design of thermo-photonic materials for passive radiative cooling. Ideal absorption coefficients and effective heat control were achieved. Wang et al. [32] introduced topological optimization formulations for the design of thermo-photonic materials, enhancing passive radiative cooling by effectively controlling energy exchanges. Shen et al. [33] studied topology optimization to improve the thermal dissipation efficiency of radiators for focal plane assemblies.

In this study, we explore a new area of topology optimization focused on lightweight laser-activated actuators, taking into account the effects of heat dissipation through convection and radiation. While previous research has addressed convection and conduction heat dissipation for actuators [34][35], to the best knowledge of the authors, there are no prior investigation into the simultaneous consideration of all three energy dissipation mechanisms (heat conduction, convection, and radiation) for actuators activated by radial energy fields such as those generated by lasers. Therefore, our research aims to comprehensively examine how radiation impacts the topological optimization design for laser-activated compliant mechanisms. We delve into the intricate role of radiation in the multiscale optimization process, offering new insights into several key areas. Firstly, we systematically analyze the influence of radiation and convection on the topological optimization design for actuators. Additionally, we thoroughly explore the spatial configuration of the design domain and the multiscale dependencies inherent in laser-activated compliant mechanisms. This exploration is elucidated through a range of case studies to provide a better understanding of the nuances of micro-scale optimization interactions within the larger multiscale optimization process.

Additionally, we examine the radiation of multi-microstructure optimization and conduct a comparative analysis, encompassing both performance considerations and additive manufacturability. Thus, the primary objective of this article is to introduce a new investigation into multiscale and multiphysics topology optimization for the creation of lightweight and compliant cellular mechanisms while considering radiation effects. Our goal is to manufacture laser-activated compliant mechanisms that are exceptionally lightweight while also addressing the influence of radiation. The structure of the article unfolds as follows: Section 2 describes the formulation of the laser-induced thermomechanical problem and details the applied homogenization technique. Section 3 provides numerical illustrations to demonstrate the effectiveness of our proposed approach. Finally, Section 4 concludes the document by summarizing our main findings and contributions.

## 2. Multiphysics Formulations for Laser-Activated Actuator

In the field of topology optimization, the material or void within a predetermined design domain $\Omega_d$ is characterized by the relative density distribution, which can be mathematically described using the following design variable $\boldsymbol{\rho}$:

$$\boldsymbol{\rho} = \begin{cases} 1 & if \ \boldsymbol{\rho} \in \Omega \\ 0 & if \ \boldsymbol{\rho} \in \Omega_d \setminus \Omega \end{cases} \qquad (1)$$

Topology optimization is a crucial procedure that requires compliance with predefined objective function constraints while achieving global extrema. The main objective of this article is the development of a formulation for laser activation for a compliant actuator. This formulation encompasses not only the energy associated with laser activation but also energy dissipation mechanisms. Subsequently, the article states the primary objective, which is to maximize displacement during actuation. As such, these topics will be addressed respectively in the following section.

### 2.1 Laser-induced material heating

Laser materials processing hinges on the intricate interplay of various physical mechanisms inherent to the process, ultimately dictating the resultant quality and performance. These outcomes are contingent upon a congruence of factors, including the specific process parameters, material selections, and the prevailing environmental conditions. In the context of this research, the foremost among these physical mechanisms is laser heating, characterized by its optical-thermal interaction[36][37]. It is crucial to understand that the quantity of excited electrons, mirroring their excitation through temperature changes, is directly proportional to the number of photons absorbed. Therefore, a laser activation regimen characterized by a heightened absorptivity, as visually depicted in Fig. 1, becomes imperative. Such enhanced absorptivity ensures efficient power conversion, a prerequisite for achieving the desired movements facilitated by the actuator.

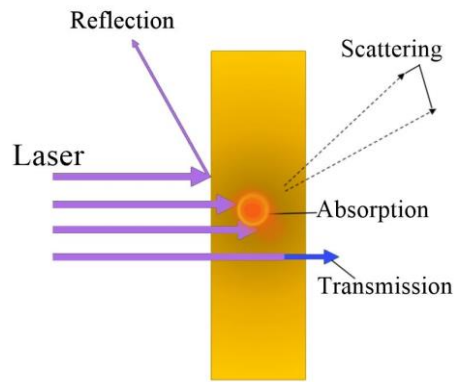

**Fig. 1** Laser interaction a medium.

In addition to the material properties, numerous parameters come into play when considering the absorption rate of a specific laser wavelength. These parameters encompass factors like surface roughness, temperature, and spatial configurations. Nonetheless, it's important to note that delving into a detailed discussion of these parameters falls outside the scope of this research. Within this section, our focus is to explore the prevalent laser heating scenarios for metals. Specifically, we examine conditions where laser irradiation primarily generates heat through absorption, without leading to the formation of any chemical compounds or the onset of intense vaporization or plasma generation near the surface. To be precise, we are referring to the scenario of low-intensity laser irradiation of metals within inert gas environments. As such, our primary focus will center on examining how the geometrical shape and size of samples, the dimensions of the irradiation spot, the energy distribution within the laser beam, variations in radiation intensity over time, and the displacement of the beam across the sample surface collectively influence the dynamics of laser heating. When it comes to the laser heating source, although the energy distribution within a cross-section of the laser beam can vary under different conditions, practicality often leads us to make certain approximations.

To start, we may assume that the laser energy distribution $I(r)$, remains consistent within a circular spot of radius $r$ characterized by an effective radius denoted as $R_s$, as shown in Eq. (2).

$$I(r) = \begin{cases} I_0 & for\ r \leq R_s \\ 0 & for\ r > R_s \end{cases} \quad (2)$$

Where $I_o$ is the laser radiation intensity. This approach is enabling for providing a reliable approximation for the highly multimode laser radiation commonly encountered in various commercial laser systems, particularly diode lasers.

However, The Gaussian distribution is regarded as the optimal pattern for laser beam distribution, particularly in the context of mono-mode laser irradiation, where typical intensity and energy density profiles are observed. Two commonly employed Gaussian distribution functions describe the laser intensity *I(r)*, and they differ based on the characteristic radius value, as illustrated in Eq (3).

$$I(r) = I_0 e^{\left(-r^2/(R_s)^2\right)} \tag{3}$$

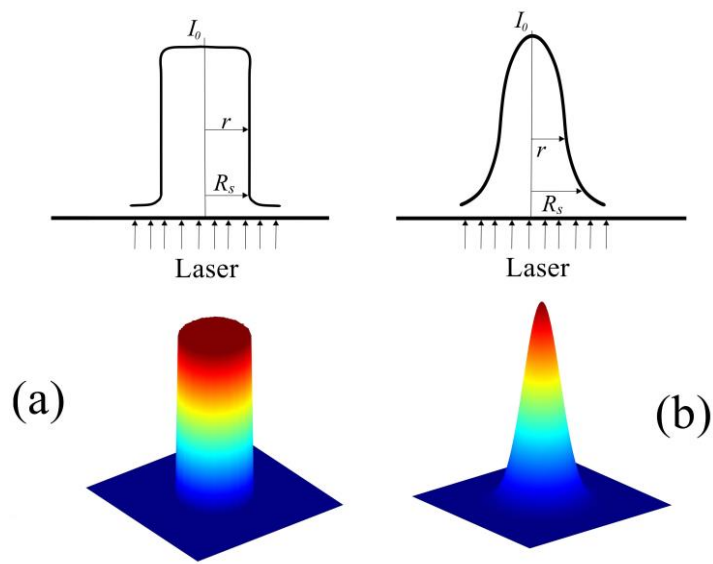

**Fig. 2** Distribution of laser beam power: (a) Uniform or Fiat intensity, (b) Gaussian intensity.

The thermal energy generated within a material by laser is intricately linked to its absorptivity; a property influenced by several key factors. These factors include the absorptivity of the incident laser's wavelength and the duration of electronic state excitation. In metallic materials, the process of energy absorption primarily involves the radiation of electrons. These excited electrons subsequently transfer their energy to the crystalline lattice through collision interactions. Moreover, it is important to emphasis that, as we delve into the laser-induced heating of materials, it's essential to consider mode distribution, a crucial factor in understanding the laser intensity profiles, before shifting our attention to heat transfer modeling and its focus on the sample's surface properties. In the context of this research, we postulate the complete absorption of the incident wavelength by the material, and the subsequent energy conversion mediated by electrons, with the further transmission of energy to the crystalline lattice through collisions occurring with negligible loss.

It is noteworthy to emphasize the thermo- spatial aspect within the interaction of laser irradiation with matter, specifically when dealing with metals, and particularly concerning the phenomenon of the anomalous skin effect [38]. The anomalous skin effect, manifesting during laser-metal interactions, exerts a profound influence upon the behavior of tire natal. This phenomenon delineates a scenario wherein the laser energy's penetration depth into the material significantly surpasses the expectations set by classical electromagnetic theory [39][40]. The phenomenon assumes particular importance when dealing with metallic substrates. In the context of an exceedingly thin metal sheet, the anomalous skin effect engenders a notable enhancement in both energy absorption and the laser beam's capacity for delving deeper into the material. Consequently, this amplification in energy deposition results in heightened thermal effects, culminating in phenomena such as increased heating and the initiation of material phase transitions, notably melting. Notably, the energy is transported further into the material than anticipated by conventional skin depth calculations. Due to the nature of the design domain and the temperature range of the working environment in the design domain which is assumed to be at room temperature (i.e., 293 Kelvin for this research), the effect of anomalous skin effect is assumed to have a negligible effect.

This disposition renders the thin metal sheet considerably more prone to structural transformations, encompassing the likes of melting, vaporization, or even ablation, contingent upon the unique parameters of the incident laser and the inherent material properties. The anomalous skin effect, therefore, has profound implications for the laser-metal interaction, especially in cases involving diminutive metal substrates, and elucidates the substantial spatial intricacies underlying this interplay.

For the heat transfer modelling, let's begin by assuming that me metallic sample is a uniform and isotropic medium. Our focus will solely be on how the surface affect the sample, primarily through changes in absorptivity. For the sake of simplicity, we will disregard volume defects and impurities. In the scenario of a semi-infinite sample, which serves as a general case from which different sample shapes are derived by imposing specific constraints. the thermal impact of the laser radiation for any given point within the system is given in Eq. (4).

$$Q_L = Q_{Cond} + Q_{Conv} + Q_{Rad} \qquad (4)$$

Where, $Q_L$ is the laser induced heat, $Q_{Cond}$, $Q_{Conv}$, and $Q_{Rad}$ are the corresponding conductions, convection, and radiation heat transfer. Laser induced heat can be written as:

$$Q_L = A_v I(r) \qquad (5)$$

Where $A_v$ is the energy absorption from radiation per unit time and unit volume of the materials. The conduction heat transfer representation as:

$$Q_{Cond} = -\kappa \nabla T \qquad (6)$$

Where $\kappa$ is the thermal conductivity, and $\nabla T$ is the temperature difference.

In the context of heat convection, we are concerned with the transfer of thermal energy between a solid surface and a fluid, whether in the form of a liquid or gas, as a consequence of the fluid's motion. This heat transfer process driven by convection is governed by Newton's Law of Cooling, encapsulated by the following equation:

$$Q_{Conv} = hA(T - T_s) \tag{7}$$

Where, $h$ is the convective heat transfer coefficient, $A$ is the surface area through which heat is convected. $T$ is the surface temperature and $T_s$ is the surrounding temperature.

Heat radiation is the transfer of heat through electromagnetic waves and is characterized by Stefan-Boltzmann's law. This law states that the rate of heat transfer through radiation is proportional to the fourth power of the absolute temperature of the radiating surface:

$$Q_{Rad} = \varepsilon \sigma A(T^4 - T_s^4) \tag{8}$$

Where, $\varepsilon$ is the emissivity of the radiating surface (a dimensionless constant between 0 and 1).

$\sigma$ is the Stefan-Boltzmann constant (approximately $5.67 \times 10^{-8}$ W/(m²·K⁴)). $A$ is the surface area. Here $T$ and $T_s$ are calculated in degree kelvin. In this research, the emissivity for radiation is considered as for perfect blackbody, which has an emissivity (ε) equal to 1.

## 2.2   Modeling Heat-Induced Thermo-Elastic Loading in Multiphysics Actuators

The phenomenon of thermal deformation is an inherent characteristic of a material, denoting its propensity to alter its volume in response to temperature gradients within its structural framework. This property is quantified in terms of the coefficient of thermal expansion, denoted as ($\boldsymbol{\alpha}_{ij}$). When a temperature gradient, symbolized by ($\Delta T$), is present, it engenders thermal loads within continuous structures as a consequence of thermal expansion. The resultant linear thermal strain, represented as ($\boldsymbol{\varepsilon}_{th}$), can be mathematically articulated as follows:

$$\boldsymbol{\varepsilon}_{th} = \boldsymbol{\alpha}_{ij} \Delta T \tag{9}$$

For an accurate representation of thermo-mechanical analyses, the incorporation of coupled-field analysis is imperative. This approach accounts for the intricate interplay and reciprocal influence between mechanical and thermal physical domains [41][42]. Two primary methodologies, namely the sequence coupling method and the direct coupling method, serve as distinct frameworks to address coupled-field analyses. These methodologies find utility in the examination of complex physical phenomena where multiple fields, such as those involving thermal and mechanical aspects, interact and interdepend. As an example, this is particularly pertinent in simulations of thermal expansion. The sequence coupling method, as one of these approaches, involves a systematic and sequential examination of the involved physical domains, adhering to the actual physical progression of the system. It employs a stepwise methodology, wherein each field is individually scrutinized, duly considering the implications of preceding analyses. In simpler terms, the analysis of one field precedes the examination of another, with the outcomes of the former analysis serving as vital input for the subsequent assessment of the latter. This approach ensures that the complex interrelations between the mechanical and thermal aspects are comprehensively accounted for during the analysis. Moreover, in our research, we have adopted the sequence coupling approach to address the thermo-elastic coupling problem. This treats the structural analysis of thermal responses as quasi-static, with thermal and elastic fields superimposed in the finite element formulation (as shown in Eq. (11)). This method allows for efficient and accurate analysis of thermo-elastic coupling while managing computational costs effectively.

$$\mathbf{KU} = \mathbf{F}_{th} \tag{10}$$

Here, $\mathbf{K}$ represents the stiffness matrix, $\mathbf{U}$ denotes the displacement, and $\mathbf{F}_{th}$ signifies the thermal-induced forces resulting from structural deformation due to temperature variance [35] which can be calculated from:

$$\mathbf{F}_{th} = \int_{\Omega} \mathbf{B}^T \mathbf{E} \boldsymbol{\alpha} \Delta \mathbf{T}^T d\Omega \tag{11}$$

Here, $\mathbf{B}$ represents the strain displacement matrix, and $\mathbf{E}$ denotes the elastic tensor of the material. The thermal gradient $\Delta \mathbf{T}$ is determined from a specialized finite element model [43], capturing heat transfer phenomena within the solid.

### 2.3 Formulations of topology optimization for heat-activated actuators

The process of creating an actuator mechanism through topology optimization centers around the quest for an optimal shape that can achieve a specified deformation pattern, as illustrated in Fig. 3. Within this context, the multiphysics of heat induced thermal strains within the elastic body demonstrates compliant demeanor, yielding a displacement denoted as ($u_{out}$) at a designated point. The primary goal of this optimization task is to ascertain the ideal arrangement of the buildings blocks that construct this elastic body (i.e., $\boldsymbol{\rho}$), with the aim of maximizing the output displacement. This must be accomplished while adhering to the equilibrium equation and adhering to a predetermined material volume constraint. In mathematical terms, the problem can be expressed as follows:

$$\max_{\boldsymbol{\rho}}: u_{out}$$

$$\text{s.t.:} \begin{cases} \mathbf{K}(\boldsymbol{\rho})\mathbf{U} = \mathbf{F}_{th} \\ \int_{\Omega_{dM}} \boldsymbol{\rho}\, d\Omega \leq v \;,\; \boldsymbol{\rho} \in (0,1] \quad \forall \boldsymbol{\rho} \in \Omega \end{cases} \quad (12)$$

Where $v$ is the volume fraction, For the porous design which is constructed of a macro design variables set $\boldsymbol{\rho}_M$ and multiscale design variables set $\boldsymbol{\rho}_m$, a separate finite element models are developed for each domain of physics, as illustrated in Fig. 3.

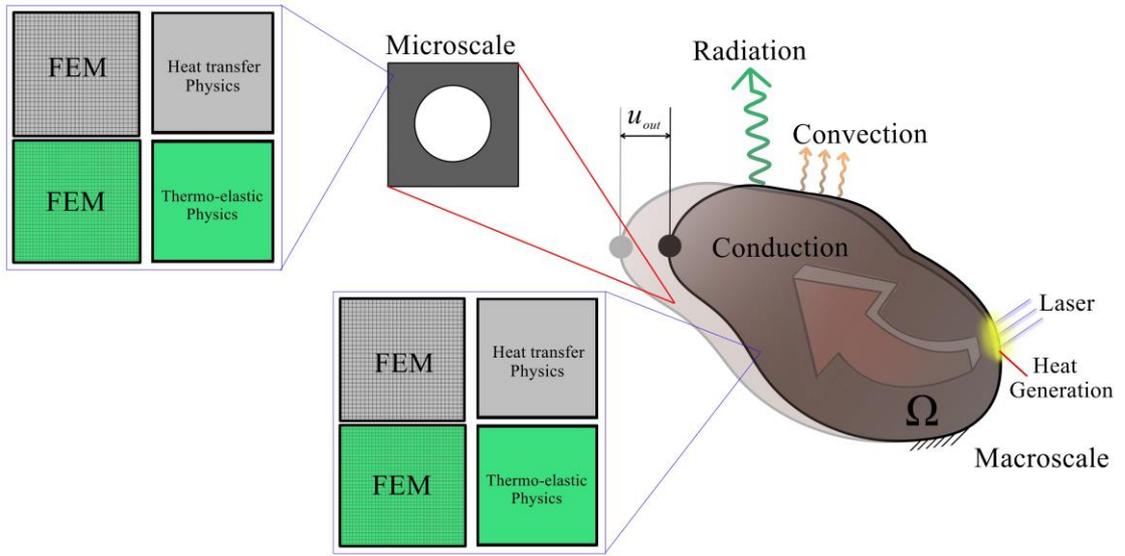

**Fig. 3.** Modeling the coupled multiphysics and multiscale of laser-activated actuators poses problem.

Therefore, we introduce the formulation of the concurrent multiscale topology optimization problem as expressed in Eq. 13.

$$\max_{\boldsymbol{\rho}_M, \boldsymbol{\rho}_m}: u_{out}$$

$$\text{s.t.:} \begin{cases} \mathbf{K}(\boldsymbol{\rho}_M, \boldsymbol{\rho}_m)\mathbf{U} = \mathbf{F}_{th} \\ \int_{\Omega_{dM}} \boldsymbol{\rho}_M\, d\Omega_{dM} \leq v_M \;,\; \boldsymbol{\rho}_M \in (0,1] \quad \forall \boldsymbol{\rho}_M \in \Omega_{dM} \\ \int_{\Omega_{dm}} \boldsymbol{\rho}_m\, d\Omega_{dm} \leq v_m \;,\; \boldsymbol{\rho}_M \in (0,1] \quad \forall \boldsymbol{\rho}_m \in \Omega_{dm} \end{cases} \quad (13)$$

Eq. 14 offers sensitivity analysis, employing the adjoint method on a coupled linear problem encompassing both heat conduction and elasticity.

$$\frac{\partial \mathbf{U}_{out}}{\partial \boldsymbol{\rho}} = \boldsymbol{\lambda}_{th}^T \left( \frac{\partial \mathbf{K}}{\partial \boldsymbol{\rho}} \mathbf{U}_{in} - \frac{\partial \mathbf{F}_{th}}{\partial \boldsymbol{\rho}} \right) + \boldsymbol{\lambda}^T \left( \frac{\partial \mathbf{K}}{\partial \boldsymbol{\rho}} \mathbf{U}_{in} \right) \qquad (14)$$

Here $\boldsymbol{\lambda}_{th}^T$ and $\boldsymbol{\lambda}^T$ are the set of adjoint vectors.

Where $\frac{\partial \mathbf{F}_{th}}{\partial \boldsymbol{\rho}}$ in for macroscale system:

$$\frac{\partial \mathbf{F}_{th}}{\partial \boldsymbol{\rho}_M} = \int_{|\Omega_M|} \mathbf{B}^T \frac{\partial \mathbf{E}^H(\boldsymbol{\rho}_M)}{\partial \boldsymbol{\rho}_M} \boldsymbol{\alpha}^H(\boldsymbol{\rho}_M) \Delta \mathbf{T} d\Omega_M \qquad (15)$$

Moreover, in terms of the micro-scale aspect:

$$\frac{\partial \mathbf{F}_{th}}{\partial \boldsymbol{\rho}_m} = \int_{|\Omega_m|} \mathbf{B}^T \frac{\partial \mathbf{E}^H(\boldsymbol{\rho}_m)}{\partial \boldsymbol{\rho}_m} \boldsymbol{\alpha}^H(\boldsymbol{\rho}_m) \Delta \mathbf{T} d\Omega_m + \int_{|\Omega_m|} \mathbf{B}^T \mathbf{E}^H \frac{\partial \boldsymbol{\alpha}^H(\boldsymbol{\rho}_m)}{\partial \boldsymbol{\rho}_m} \Delta \mathbf{T} d\Omega_m \qquad (16)$$

while $\frac{\partial \mathbf{K}}{\partial \boldsymbol{\rho}}$ in term of macroscale is

$$\frac{\partial \mathbf{K}}{\partial \boldsymbol{\rho}_M} = \int_{|\Omega_M|} \mathbf{B}^T \frac{\partial \mathbf{E}^H(\boldsymbol{\rho}_M)}{\partial \boldsymbol{\rho}_M} \mathbf{B} \, d\Omega_M \qquad (17)$$

And in term of multiscale design variable

$$\frac{\partial \mathbf{K}}{\partial \boldsymbol{\rho}_m} = \int_{|\Omega_m|} \mathbf{B}^T \frac{\partial \mathbf{E}^H(\boldsymbol{\rho}_m)}{\partial \boldsymbol{\rho}_m} \mathbf{B} \, d\Omega_m \qquad (18)$$

The derivative of the homogenized material's elastic tensor with respect to the micro scale design variable $\frac{\partial \mathbf{E}^{H}(\boldsymbol{\rho}_{m})}{\partial \boldsymbol{\rho}_{m}}$ can be calculated as [44]:

$$\frac{\partial \mathbf{E}^{H}(\boldsymbol{\rho}_{m})}{\partial \boldsymbol{\rho}_{m}} = \frac{p}{|\Omega_{m}|} \int_{\Omega_{m}} \left(\boldsymbol{\rho}_{m}^{p-1}\right) \mathbf{E}_{ijqp}^{0} \left(\boldsymbol{\varepsilon}_{qp}^{0(kl)} - \boldsymbol{\varepsilon}_{qp}^{*(kl)}\right) d\Omega_{m} \tag{19}$$

Here $p$ represents the penalization factor of the corresponding multiscale design vector. For further insights into the homogenization method and its implementation in concurrent multiscale topology optimization, the interested readers may refer to [44]. The derivative of the homogenized material's thermal expansion vector with respect to micro design variable $\frac{\partial \boldsymbol{\alpha}^{H}(\boldsymbol{\rho}_{m})}{\partial \boldsymbol{\rho}_{m}}$ can be represented as [45]:

$$\frac{\partial \boldsymbol{\alpha}^{H}(\boldsymbol{\rho}_{m})}{\partial \boldsymbol{\rho}_{m}} = \frac{1}{\mathbf{E}^{H}} \left( \frac{\partial \boldsymbol{\beta}^{H}(\boldsymbol{\rho}_{m})}{\partial \boldsymbol{\rho}_{m}} + \boldsymbol{\alpha}^{H}(\boldsymbol{\rho}_{m}) \frac{\partial \mathbf{E}^{H}(\boldsymbol{\rho}_{m})}{\partial \boldsymbol{\rho}_{m}} \right) \tag{20}$$

Where $\boldsymbol{\beta}^{H}$ is the homogenized thermal stress tensor. Here the term $\frac{\partial \boldsymbol{\beta}^{H}(\boldsymbol{\rho}_{m})}{\partial \boldsymbol{\rho}_{m}}$ can be determined as:

$$\begin{aligned}\frac{\partial \boldsymbol{\beta}^{H}(\boldsymbol{\rho}_{m})}{\partial \boldsymbol{\rho}_{m}} &= \frac{p}{|\Omega_{m}|} \int_{\Omega_{m}} \left(\boldsymbol{\varepsilon}_{qp}^{0(kl)} - \boldsymbol{\varepsilon}_{qp}^{*(kl)}\right)^{T} \left(\boldsymbol{\rho}_{m}^{p-1}\right) \mathbf{E}_{ijqp}^{0} \left(\boldsymbol{\rho}_{m}^{p} \boldsymbol{\alpha} - \boldsymbol{\varepsilon}^{th}\right) d\Omega_{m} + \\ &\quad \frac{p}{|\Omega_{m}|} \int_{\Omega_{m}} \left(\boldsymbol{\varepsilon}_{qp}^{0(kl)} - \boldsymbol{\varepsilon}_{qp}^{*(kl)}\right)^{T} \boldsymbol{\rho}_{m}^{p} \mathbf{E}_{ijqp}^{0} \left(\boldsymbol{\rho}_{m}^{p-1}\right) \boldsymbol{\alpha} \, d\Omega_{m}\end{aligned} \tag{21}$$

Thus, the multiscale system model is shown in Fig. 4.

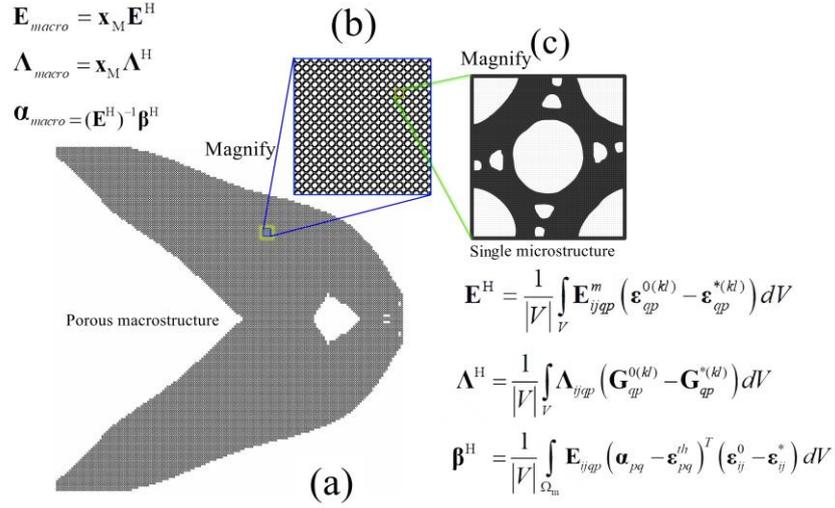

**Fig. 4.** Multiscale design challenge: (a) macroscopic structure; (b) periodically arranged cellular structure; (c) microscopic structure.

This study employs a comprehensive methodology involving two systems: macroscale and multiscale. Each scale is discretized using specific finite elements corresponding to unique design variables, defining the optimized state of solid materials or voids within respective design domains. The optimization unfolds through sequential steps ensuring precision and reliability in the final structural configuration. Step 1 initiates by defining pivotal design parameters, including discretization size, desired volume fraction, and material properties such as thermal conductivity and mechanical constitutive matrix. Step 2 treats each element within the macro design domain as a unit cell, facilitating finite element analysis within a thermo-elastic coupled system. Multiscale discretization follows, employing homogenization theory to compute effective macroscopic properties of the Representative Volume Element (RVE). Step 3 determines effective properties before initiating topology optimization at the macroscale. Step 4 involves constructing the objective function and evaluating displacement within the design domain. Step 5 includes sensitivity analysis and application of optimality criteria, followed by design domain updates. Step 6 scrutinizes design variables related to constraint functions to ensure weight minimization. Step 7 iteratively repeats steps 2 to 7 until achieving the desired configuration with optimal efficiency [43]. Fig. 5 illustrates the flowchart detailing the concurrent multiscale and multiphysics topology optimization methodology for maximizing actuation displacement in a laser-activated displacement actuator.

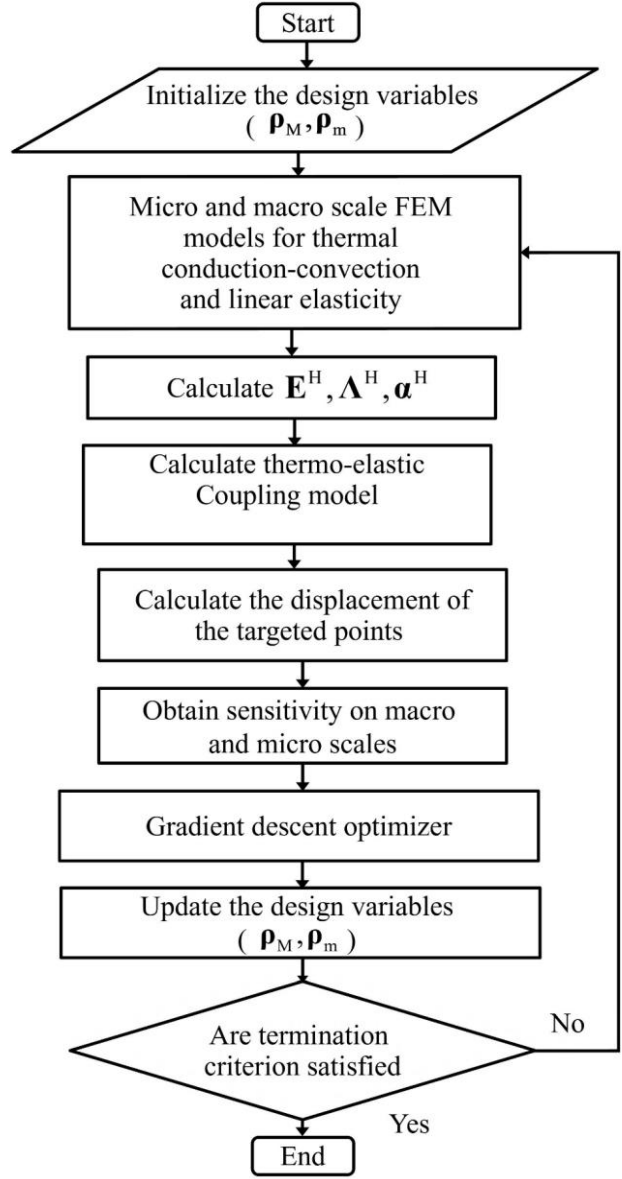

**Fig. 5.** Flowchart for optimizing mechanical displacement in a laser activated actutaor using concurrent multiscale and multiphysics topology optimization.

## 3. Numerical Investigation

In this section, we delve deeply into thermoelastic formulations within multiphysics multiscale scenarios, utilizing a range of numerical examples. The investigation is performed into three sub-sections, each focusing on specific aspects of the optimization process. Firstly, we examine a laser-activated actuator considering heat dissipation through heat conduction. Next, we explore convection and radiation effects. Finally, we discuss a study addressing the spatial influence on topology optimization as well as the multiscale optimization dependency on microstructure.

**3.1 Laser-activated displacement actuator optimization under the consideration of energy dissipation via heat conduction**

In this section, our focus is on designing laser-activated displacement actuators, accounting for heat conduction in solid and porous examples. The aim is to engineer actuators capable of converting thermal energy into linear motion. We examine a single scale 2D macro design domain, thermally insulated from top and bottom boundaries (Fig. 6(a)). Bilinear finite element discretization is used for both elastic and solid heat transfer physics domains (Fig. 6(b)). This approach facilitates a comprehensive exploration of the actuator's behavior under thermal loading conditions. The heat source in this study is modeled as a laser pulse characterized by a Gaussian distribution (i.e., $TEM_{00}$). This laser pulse induces a heat wave that propagates radially from the laser irradiation point. Therefore, in the rectangular design domain illustrated in Fig. 6(a), temperatures in areas close to the fixed point are lower than those at the center of the left side. For our numerical analysis, we assume an initial temperature of 50 degrees Celsius (corresponding to 323.15 Kelvin at fixed points) due to the heat generated by laser irradiation. This temperature increases parabolically until reaching 100 degrees Celsius (373.15 Kelvin) at the center of the left side of the design domain. We chose to express all temperatures in Kelvin to ensure consistency with the subsequent consideration of thermal radiation. In this study, we focus solely on the heat transport phenomenon within a solid medium. Heat propagation is limited to occur within the solid, and thermal convection and radiation will be considered in the forthcoming analyses.

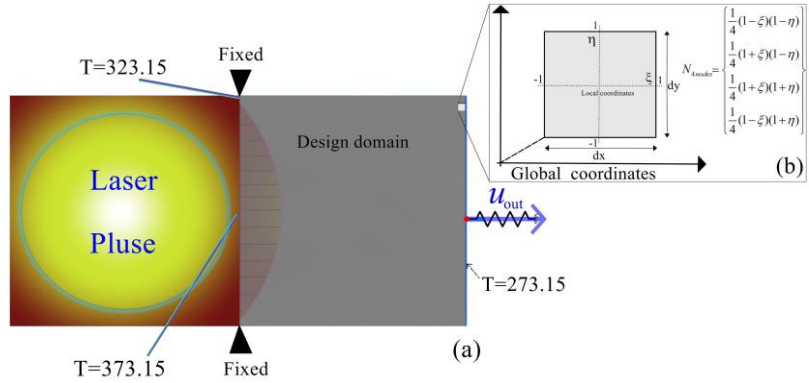

**Fig. 6** Micro-scale model depicting the laser-triggered displacement actuator in a single-scale configuration.

In this study, meticulous consideration is given to the physical properties of the materials involved in all numerical simulations. Parameters such as Young's modulus, thermal conductivity, and expansion coefficients are assumed to have a unit value for simplification purposes, facilitating focused analysis of the phenomenon under study. During the topology optimization process, void elements are characterized by infinitesimal values rather than zero. This deliberate choice minimizes numerical errors, thereby enhancing the precision and reliability of simulation results, leading to a more accurate representation of the studied phenomenon [34][46][47]. Each model in this study uses a mesh of 100 by 100 µm in the x and y directions, with bilinear elements. Each element is assigned an area of 1 µm². The optimization process is initially performed at a single scale, specifically at the macro-scale level. Two distinct volume fractions are considered: (a) a volume fraction of 0.5 corresponding to a 50% weight reduction, and (b) a volume fraction of 0.25 corresponding to a 75% weight reduction. The results of the single-scale topology optimization of the laser-activated actuator under the sole physics of thermal conduction are illustrated in Fig. 7.

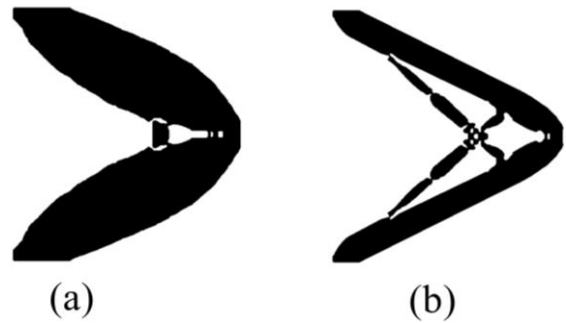

**Fig. 7** Macroscale topology optimization results for the laser-activated actuator, presenting (a) 50% weight reduction and (b) 75% weight reduction.

Let's examine the design of porous actuators, where porous structures demonstrate exceptional performance characterized by a substantial weight reduction. While our investigation extends in the context of multiscale optimization, it is necessary to construct the microstructure using a dedicated finite element model. Following the theory of numerical homogenization, we systematically address the microstructure by incorporating periodic boundary conditions. This methodology ensures the accurate representation of the micro-scale system behavior within the overall multiscale framework, facilitating in-depth analysis of interactions and optimizations at both macro and micro levels [48]. Therefore, an identical finite element mesh discretization to the single-scale case was chosen for the micro-scale. This mesh configuration provides adequate resolution of the microstructure while simultaneously ensuring computational efficiency. The decision to use this mesh size is strategic, striking a balance between capturing intricate details of the microstructure and maintaining computational feasibility. It is worth to mention that the homogenization process and sensitivity analysis of the microstructure require substantial computational power due to the need to address extensive sets of equations, describing the material behavior at the microscopic scale, as illustrated in Fig. 8.

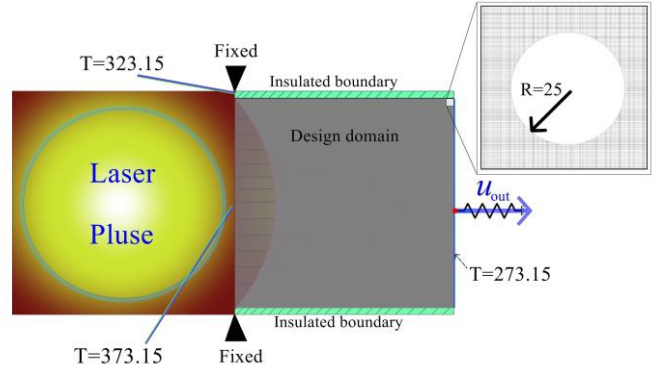

**Fig. 8** Design domain for multiscale laser activated actuator.

In this investigation, our approach involved establishing an initial design domain for the microstructure, characterized by a central sub-domain of design in a circular configuration with a diameter of 50 elements. The extent of this central sub-domain design was deliberately set at an infinitesimal initial level.

The significance pertaining to the remainder of the original microstructure design domain persists. This particular microstructure arrangement was selected for two primary purposes: firstly, to streamline the attainment of an optimal design with minimized computational expenses, and secondly, to alleviate instability arising from numerical inaccuracies. The overarching objective of multiscale optimization aimed to achieve a 50% reduction in weight across both macro and micro levels, ultimately culminating in a total volume fraction of 0.25, corresponding to a 75% reduction in weight. This premise stems from the assumption that each element possesses a volume equivalent to that of the micro-scale volume. By concurrently minimizing weights at both micro and macro scales, substantial volume reduction can be attained. The outcomes of the optimization procedures at macro and micro scales are depicted in Fig. 9.

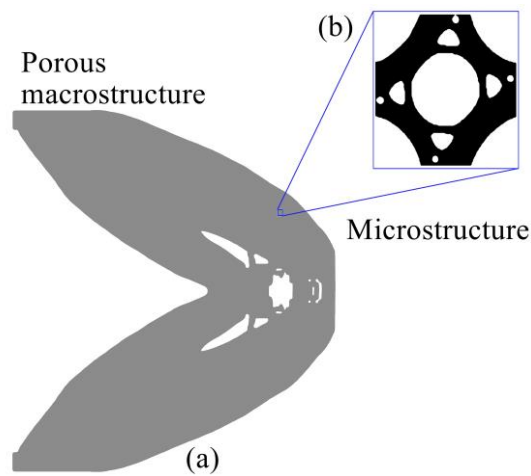

**Fig. 9** Multiscale topology optimization of porous actuator (a) macrostructure, (b) microstructure unit cell.

The graphical representation of performance measures in µm for single and multiscale optimized results is described in Fig. 10. In the realm of physical implications, a design featuring an increased membrane thickness inherently exhibits enhanced resistance to motion as a compliant mechanism, attributable to its increased stiffness and concomitant reduction in compliance. Conversely, a design incorporating a finer linkage can achieve an extended range for an equivalent actuation field. This distinction is perceptible in the results, depicting the maximum displacement obtained for a 50% weight reduction compared to a 75% weight reduction.

Furthermore, the concurrently optimized multiscale porous displacement actuator, enabling a 75% weight reduction, exhibits superior performance compared to a proportionate weight reduction in the single-scale case. It demonstrates a remarkable improvement of 45.5% compared to the 50% weight reduction and a substantial enhancement of 33.3% compared to the single-scale case with a 75% weight reduction. The increased freedom in the microscopic design domain allows the concurrent optimization algorithm to distribute materials more efficiently for the heat-activated porous displacement actuator. This influence extends to both the macro arrangement and the overall layout, demonstrating the effectiveness of the multiscale approach in enhancing the performance of the laser-activated actuator.

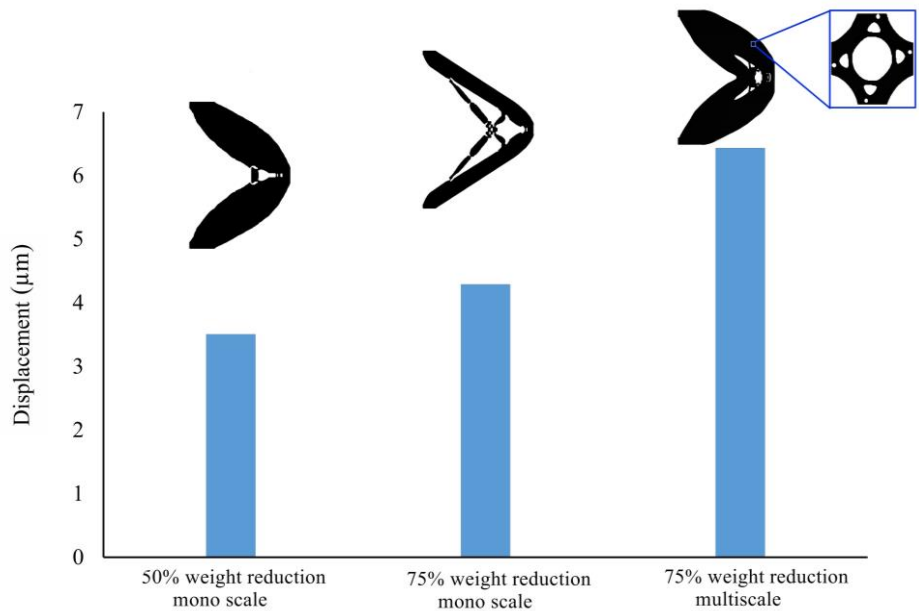

**Fig. 10** Efficiency of the solid with a 50% weight reduction, the solid with a 75% weight reduction, and the porous actuator with a 75% weight reduction.

We expand our exploration of porous displacement actuator production by undertaking sequential optimizations at both macro and micro scales, as opposed to simultaneous multiscale optimization. Our primary goal is to discern variations, particularly considering the interdependent responses of the two systems throughout the optimization process. Consequently, we perform the optimization of the same configuration at the micro-scale as depicted in Fig. 8. However, we perform the optimization exclusively at the micrometer scale, utilizing the macro-scale optimized design with a 50% weight reduction, as illustrated in Fig. 7 (a). The results are depicted in Fig. 11. The outcomes indicate that successive topological optimization yields a microstructure exhibiting similarities with maximizing the shear modulus in inverse homogenization topological optimization for microstructure design. In terms of performance, it outperforms the single-scale optimized casing, allowing for a 75% weight reduction. However, its performance is insufficient compared to the case of simultaneous multiscale optimization (as shown in Fig. 11). This discrepancy can be attributed to the mutual update of micro and macro scales at each iteration, providing the design process with the flexibility needed to achieve designs with increased agility in terms of movement.

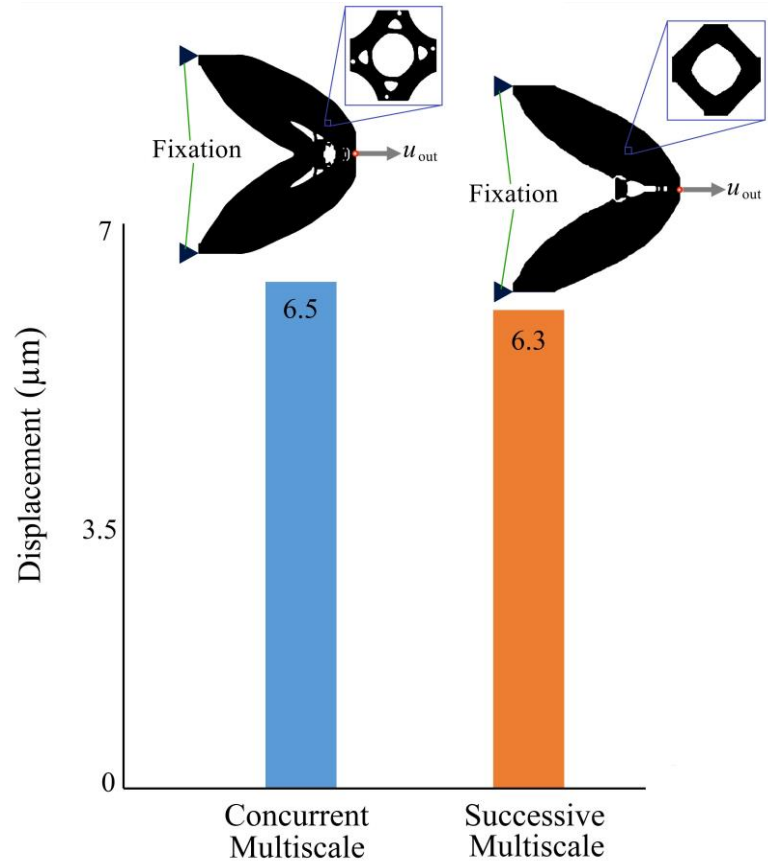

**Fig. 11** Comparative results of successive and concurrent multiscale topology optimizations for a thermally conductive actuator.

### 3.2 Optimization of the laser-activated displacement actuator considering thermal dissipation with conduction and convection

This study extends its scope to explore the influence of thermal convection on a single-scale design domain similar to that represented in Fig. 6, encompassing weight reductions of 50% and 75%. The impact of increased thermal convection on optimized cases is meticulously examined by conducting optimizations across a spectrum of thermal convection coefficients. The ambient temperature is consistently set at 283.15 Kelvin. It is worth emphasizing the practice advocated by many researchers in the field of topological optimization, recommending the use of dimensionless studies. As such, numerous research studies have employed material constants set to unity, rationalizing the study and facilitating the application of recommended objective functions and gradient-based optimizers. This standardized approach not only enhances the understanding of adjoint analysis efficiency but also simplifies the exploration of complex problems through dimensional analysis. In our dimensionless study, we start with a small value for the convection rate. This cautious approach stems from the fact that both thermal conduction and dilation coefficients are set to a value of 1 for the base material in all design cases. Therefore, even a minor adjustment of surface convection boundaries can have a significant impact on topological optimization results. This highlights the system's sensitivity to variations in convection rates, emphasizing the need for nuanced exploration of thermal convection effects in the optimization process. The results of this investigation, detailed in Table 1, reveal a notable response of the design domain to increased thermal convection. This response is manifested in the development of extensions emanating from the main design body. Extending the exploration of convection to multiscale optimization, as presented in Table 2, highlights a parallel trend to single-scale optimization results. Alongside increased thermal convection, the macro-scale of the multiscale optimized casing exhibits an extension of its main design body through the addition of extensions. This adaptive behavior in response to increased thermal convection emphasizes the role of convection in determining design resilience. Furthermore, the results unveil a distinctive pattern along the design boundaries, where spike-shaped formations emerge and extend from the main body. These formations gain significance with intensified thermal convection. The macro-scale design, influenced by increased heat transfer due to convection, incorporates perpendicular links within its structure, causing a corresponding adaptation in the microstructure arrangement. As indicated in Table 2, the micrometer-scale responds to this phenomenon by increasing the shear modulus and decreasing the elastic modulus.

The Young's modulus in the x and y directions. This adaptive strategy aims to enhance resilience and elasticity against thermally induced actuation while simultaneously reinforcing resistance to torsion that could impede linear movement in the x-direction, the maximization objective of the optimization process. Moreover, the solid thermal conductivity tensor implies a concerted effort within the microstructure to reduce heat transfer to the newly formed extensions, further illustrating the complex interaction between design adjustments and thermal convection dynamics.

Table 1 Topologically optimized results of laser-activated displacement actuator at the macro scale with varying convection coefficients.

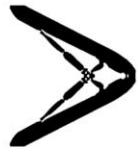

Table 2 Variations in thermal convection coefficients in the multiscale design of a laser-activated porous displacement actuator.

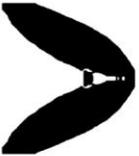

**3.3 Optimization of the laser-activated displacement actuator considering thermal dissipation with conduction, convection, and radiation**

In this section, we expand our investigation to include the impact of coupled thermal radiation with thermal convection. The problem assumes radiative non-uniformity at the boundaries of the designed structure. Additionally, we consider the radiation to be entirely diffusive. The analysis examines the radiation emission from the surfaces of the designed structure to its environment. We assume that the boundary layer effect and material properties of the structure have a negligible influence. The consideration of polarization is omitted in this study. Furthermore, the environment is assumed to have a unit refractive index. The analysis focuses on a compact structure in space, where the contribution of reflected energy at the surface of the structure is deemed negligible. Let's commence our examination by focusing on the single scale, similar to that illustrated in Section 3.1.

Utilizing a design domain and assumptions analogous to those applied in Fig. 6, where convection is considered at a given rate and radiation is assumed to occur only on the outer boundaries of the topologically optimized structure, as highlighted earlier in this section. Furthermore, the volumetric fraction condition is maintained at 0.5 (i.e., a 50% weight reduction). The resulting design is depicted as the first case in Table 3.

Table 3 Single-scale macro design of a laser-activated displacement actuator under convection and radiation with various spatial configurations.

| Case | Macroscale design |
|---|---|
| 1 | 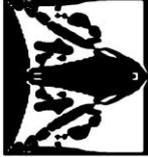 |
| 2 | 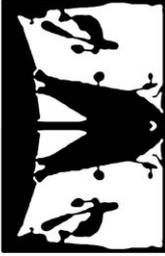 |
| 3 | 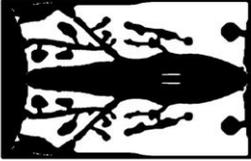 |

The results indicate a tendency to construct radiation dissipation surfaces oriented towards space while maintaining a relatively extensive surface to uphold a balance of thermal dissipation through convection. For a more in-depth exploration of the single configuration relevant to topological optimization for a laser-activated actuator, two cases involving elongated and extruded design domains with a volumetric fraction condition of 0.5, each possessing identical areas, were investigated. The initial design domains for these two cases are depicted in Fig. 12. Case (a) features dimensions of 125 µm in height and 80 µm in width, while the second case has dimensions of 80 µm in height and 125 µm in width, as shown in Fig. 12 (b). The topologically optimized results for cases (a) and (b) in Fig. 12 are presented in Table 3 for cases 2 and 3, respectively.

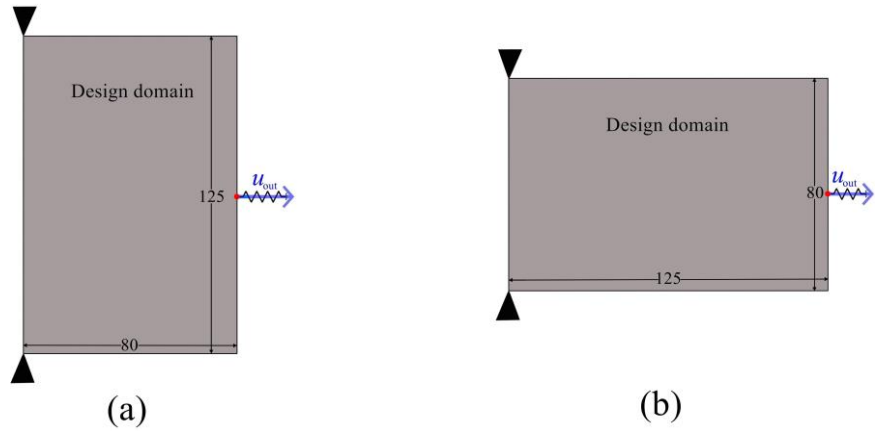

**Fig. 12** Comparison of initial design domains for topological optimization: (a) Elongated design with a height of 125 µm and a width of 80 µm, and (b) Extruded design with a height of 80 µm and a width of 125 µm.

On the other hand, our exploration revolves around the concept of creating a porous actuator while considering the dissipation of energy through conduction, convection, and radiation mechanisms. Additionally, we conducted a fundamental comparative study by delving into the multiscale topological optimization design, as illustrated in Fig. 8. In this context, we examined the effects of radiation at the macro scale, based on the previous example in this section. The primary thermal dissipation within porous media is exclusively attributed to conduction, minimizing both radiation leakage and convection. The design process unfolds simultaneously, ensuring a dynamic exchange of influences between the macro and micro design domains, ultimately aiming for superior results. The outcomes of this simultaneous multiscale topological optimization design are depicted in Fig. 13.

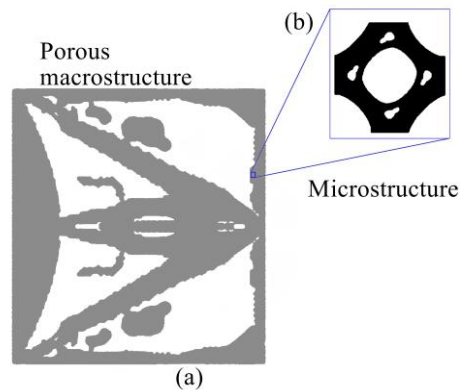

**Fig. 13** Multiscale topological optimization of a laser-activated displacement actuator under convection and radiation: (a) macrostructure, (b) unit cell of microstructure.

It is important to mention that the study of the convection rate in the context of the porous laser-activated actuator is of significant importance, providing crucial insights into the thermal dynamics and overall system efficiency, especially considering the realization of the design for various applications. In this regard, a study was conducted, comparing three different convection rates to discern their impact on the simultaneous multiscale optimization design of the actuator. The investigation involved increasing the convection rates while keeping another parameter constant. The results of this comparative study revealed a notable trend towards an increase in the surface of the convective mechanism, particularly in terms of macro-scale extensions, as presented in Table 4.

Table 4  Comparative analysis of the multiscale design of a porous laser-activated actuator with a micro-scale of the initial central circle design sub-domain, under different convection rates.

| Convcetion Rate | Macroscale design | Microscale design |
|---|---|---|
| 1.00E-07 | 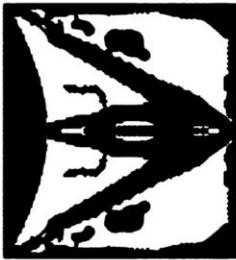 | 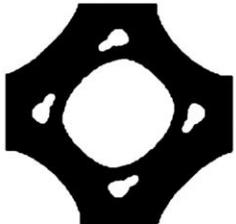 |
| 2.00E-07 | 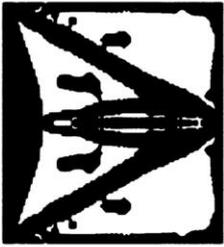 | 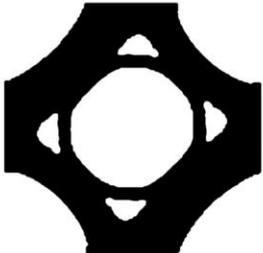 |
| 3.00E-07 | 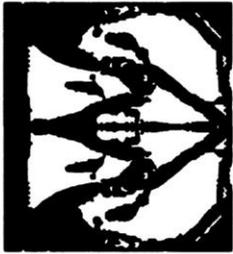 | 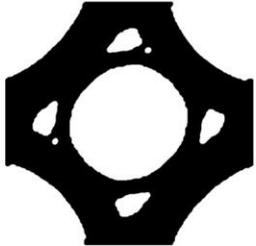 |

The observed trend towards an enlarged surface as convection rates vary can be scientifically explained by considering the thermal dissipation mechanisms within the porous structure. At lower convection rates, the movement of thermal energy relies primarily on conduction, as thermal dissipation through convection is relatively limited. With an increase in convection rates, the dynamic nature of fluid flow enhances heat transfer through convection, thereby promoting more efficient thermal energy exchange. This increased fluid movement facilitates more effective heat dissipation, resulting in an increased effective surface for thermal exchange. The simultaneous process of multiscale topology optimization, driven by the complex interaction between macro and micro-scale effects, prompted an investigation into the dependence of the macrostructure on microstructure systems [34]. To explore this relationship, multi-microstructures were introduced and simultaneously optimized with the macrostructure in several cases. This investigation involved varying the micrometer-scale. Initial design domain, categorized into three main groups, each comprising four distinct subgroups. In the first category (as shown in Fig. 14 (b)), the initial subgroup featured a central circular design subdomain. The second subgroup of the first category positioned an eccentric circle in the top-left corner (as shown in Fig. 14 (c)), while the third subgroup involved three distributed circular holes (as shown in Fig. 14 (d)). The fourth subgroup of the first category included a design domain with four circular holes (as shown in Fig. 14 (e)). This study was conducted at various convection rates. Specifically considering the micrometer-scale with a central circular hole (the first subgroup of the first category), already presented in Table 4, the investigation extended to other subgroups, detailed in Tables 5 to 7. These tables offer a detailed examination of the diverse sub-design domains' impacts on the optimization process and their resulting effects on the design of the laser-activated actuator. The systematic exploration of variations at the micrometer scale provides valuable insights into holistic design considerations for achieving superior results in the simultaneous multiscale topology optimization of porous actuators.

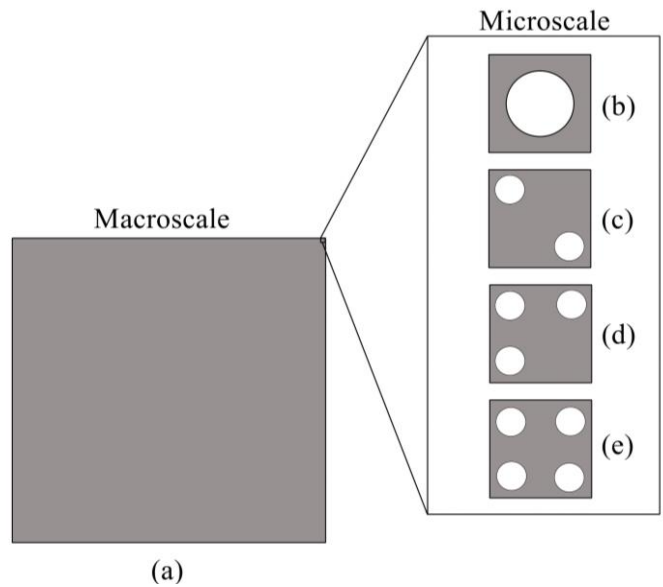

**Fig. 14** The category 1 design domains with micro-scale subgroups (1-4) in a simultaneous multiscale topology optimization for a laser-activated actuator.

Table 5 Comparison of multiscale designs for laser-activated porous actuators featuring a micro-scale with an initially positioned eccentric circle in the upper left corner, assessed under different convection rates.

| Convcetion Rate | Macroscale design | Microscale design |
|---|---|---|
| 1.00E-07 | 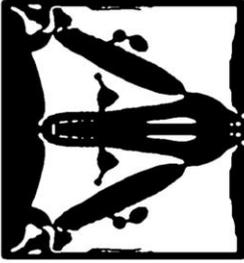 | 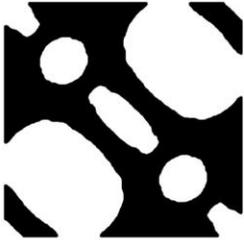 |
| 2.00E-07 | 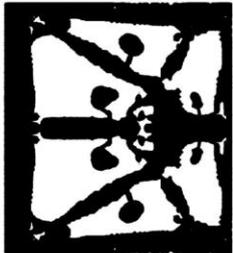 | 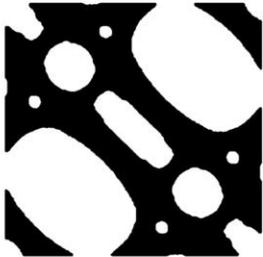 |
| 3.00E-07 | 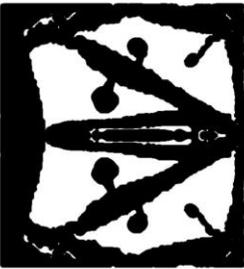 | 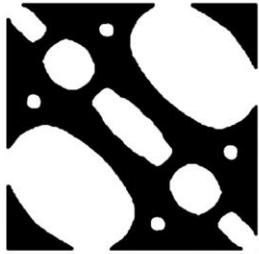 |

Table 6  Comparison of multiscale designs for laser-activated porous actuators featuring a micro-scale with three initial circular holes, evaluated under various convection rates.

| Convcetion Rate | Macroscale design | Microscale design |
|---|---|---|
| 1.00E-07 | 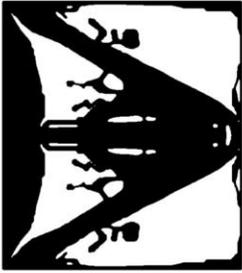 | 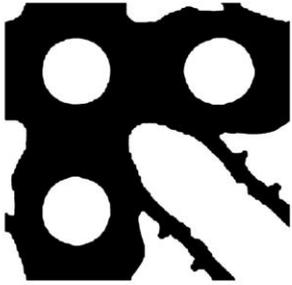 |
| 2.00E-07 | 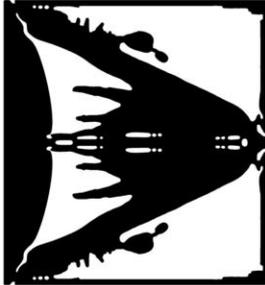 | 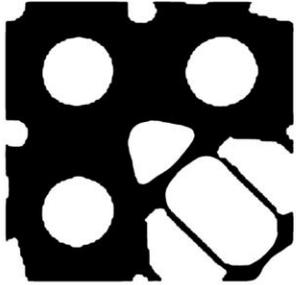 |
| 3.00E-07 | 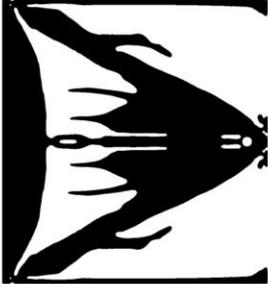 | 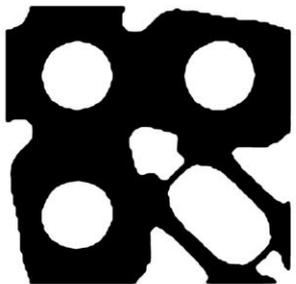 |

Table 7 Comparison of multiscale designs for laser-activated porous actuators featuring a micro-scale with four initial circular holes, evaluated under various convection rates.

| Convcetion Rate | Macroscale design | Microscale design |
|---|---|---|
| 1.00E-07 | 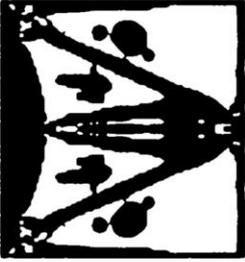 | 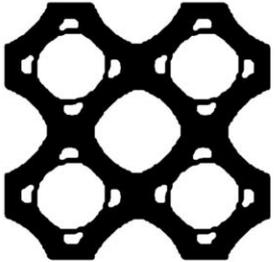 |
| 2.00E-07 | 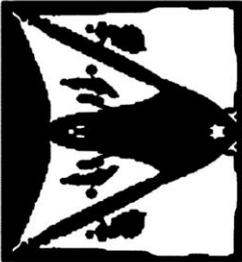 | 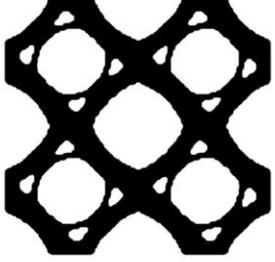 |
| 3.00E-07 | 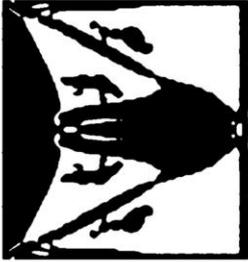 | 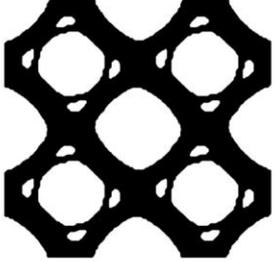 |

In the second category (Fig. 15 (b)), emphasis was placed on a central square design sub-domain for the initial subgroup. The second subgroup (Fig. 15 (c)) featured an eccentric square positioned in the upper-left corner, while the third subgroup (Fig. 15 (d)) incorporated three square holes. The fourth subgroup of the second category included a design domain with four square holes (Fig. 15 (e)). In-depth analyses of all subgroups are documented in Tables 8 to 11, elucidating their contributions to the optimization process.

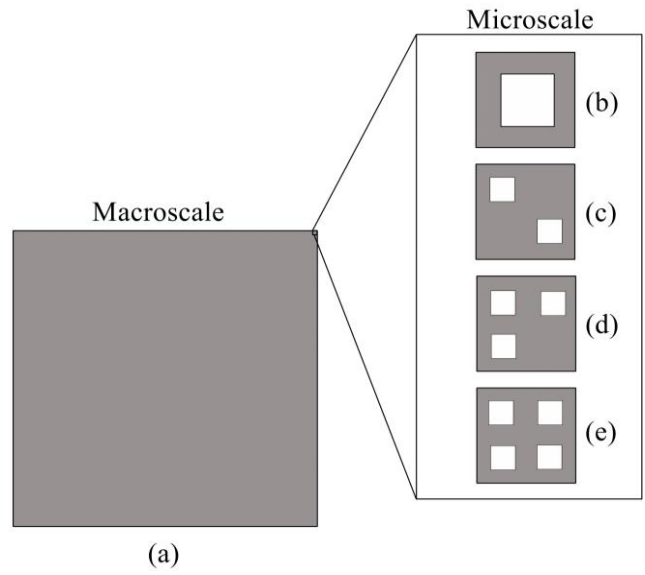

**Fig. 15** Category 2 design domains with micro-scale subgroups (1-4) in a simultaneous multiscale topological optimization achieved a laser-activated actuator.

Table 8 Comparative analysis of the multiscale design of a laser-activated porous actuator with a microscopic design sub-domain of the central square, under different convection rates.

| Convcetion Rate | Macroscale design | Microscale design |
|---|---|---|
| 1.00E-07 | 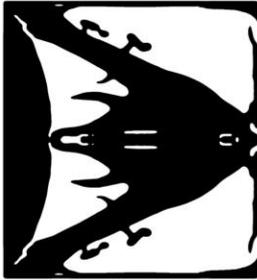 | 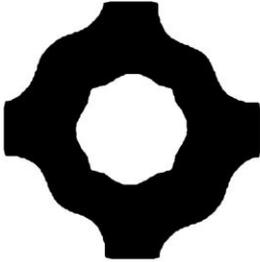 |
| 2.00E-07 | 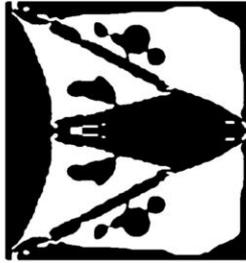 | 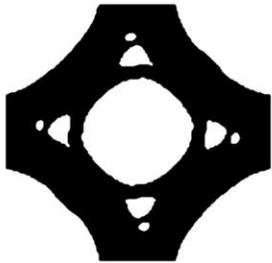 |
| 3.00E-07 | 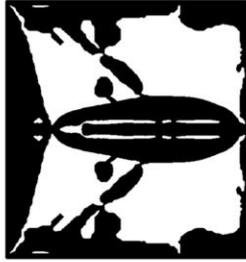 | 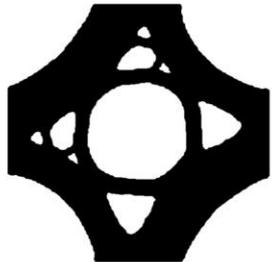 |

Table 9 Comparison of multiscale designs for laser-activated porous actuators featuring a micro-scale with an eccentric square initially positioned in the top-left corner, evaluated under different convection rates.

| Convcetion Rate | Macroscale design | Microscale design |
|---|---|---|
| 1.00E-07 | 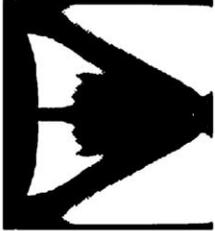 | 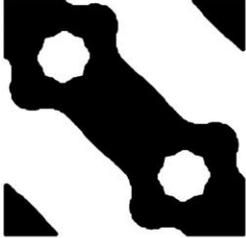 |
| 2.00E-07 | 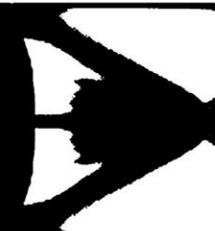 | 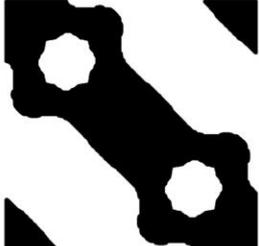 |
| 3.00E-07 | 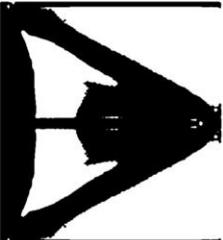 | 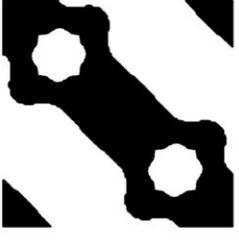 |

Table 10  Comparative analysis of multiscale designs for laser-activated porous actuators featuring a micro-scale with initial three-square holes, evaluated at different convection rates.

| Convcetion Rate | Macroscale design | Microscale design |
|---|---|---|
| 1.00E-07 | 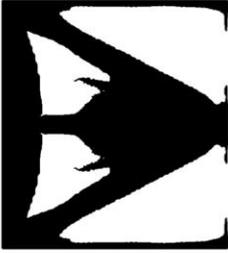 | 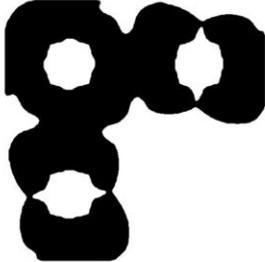 |
| 2.00E-07 | 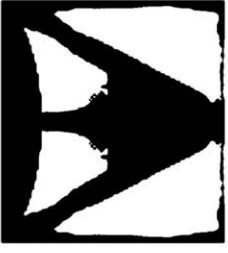 | 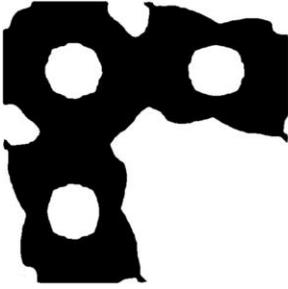 |
| 3.00E-07 | 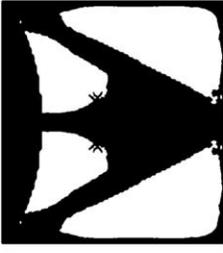 | 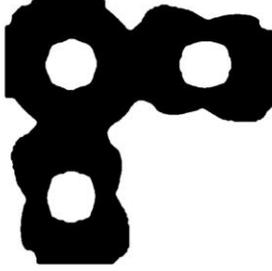 |

Table 11 Comparative analysis of multiscale designs for laser-activated porous actuators featuring a microscale with three or four initial holes, evaluated at different convection rates.

| Convcetion Rate | Macroscale design | Microscale design |
|---|---|---|
| 1.00E-07 | 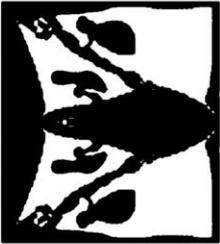 | 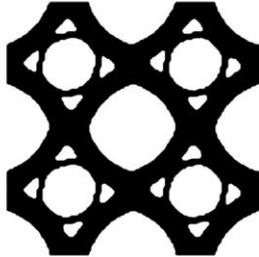 |
| 2.00E-07 | 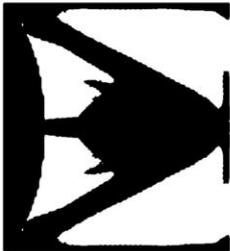 | 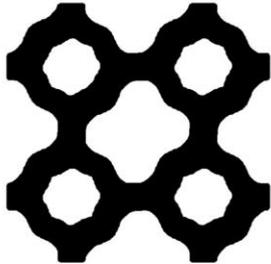 |
| 3.00E-07 | 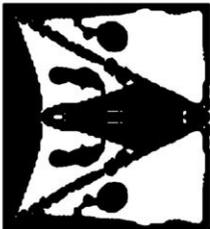 | 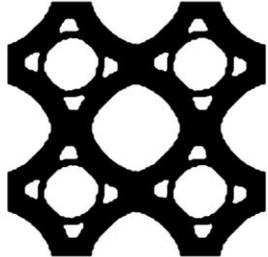 |

In the third category (Fig. 16 (b)), emphasis was placed on a specific sub-design area of the central diamond for the initial subgroup. The second subgroup (Fig. 16 (c)) featured an eccentric diamond positioned in the upper left corner, while the third subgroup (Fig. 16 (d)) incorporated three diamond-shaped voids. The fourth subgroup of the second category presented a design area with four diamond-shaped voids (Fig. 16 (e)). Further analyses of all subgroups are detailed in Tables 12 to 15, explaining their contribution to the optimization process.

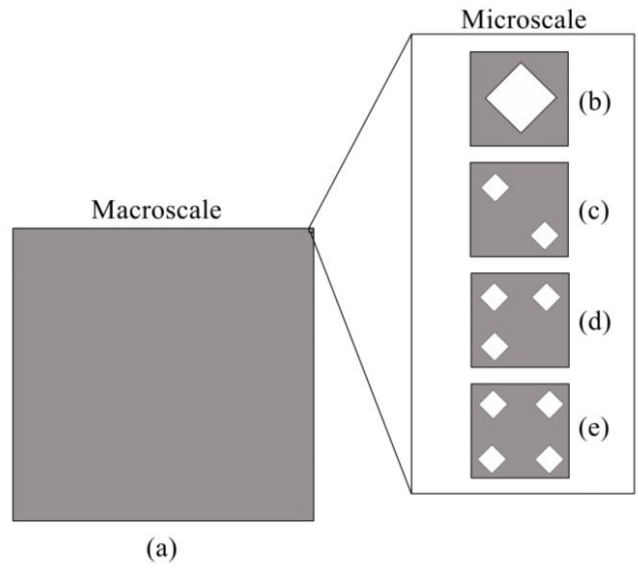

**Fig. 16** Category 3 design domains with micro-scale subgroups (1-4) in a simultaneous multiscale topological optimization obtained a laser-activated actuator.

Table 12 Comparative analysis of the multiscale design of a laser-activated porous actuator with a micro-scale of the initial sub-design domain in the central diamond, under different convection rates.

| Convcetion Rate | Macroscale design | Microscale design |
|---|---|---|
| 1.00E-07 | 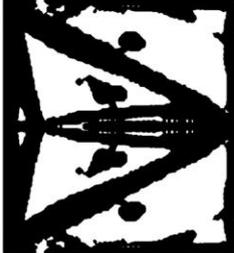 | 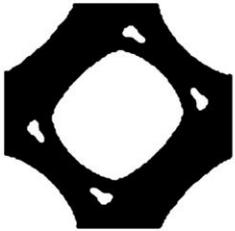 |
| 2.00E-07 | 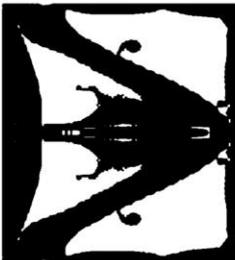 | 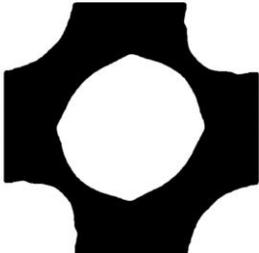 |
| 3.00E-07 | 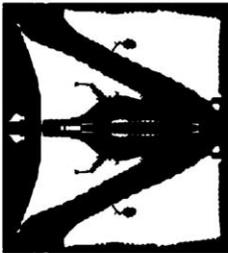 | 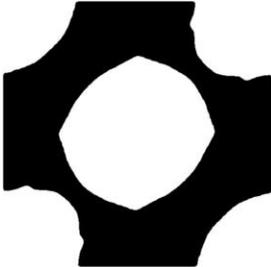 |

Table 13 Comparison of multiscale designs for laser-activated porous actuators incorporating a micro-scale with an initial eccentric diamond positioned in the upper left corner, evaluated under different convection rates.

| Convection Rate | Macroscale design | Microscale design |
|---|---|---|
| 1.00E-07 | 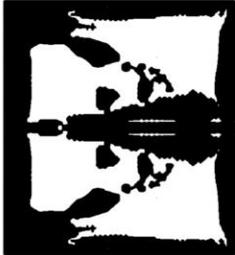 | 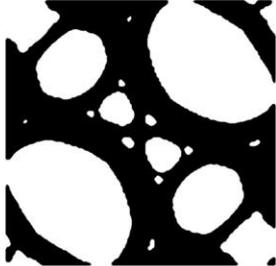 |
| 2.00E-07 | 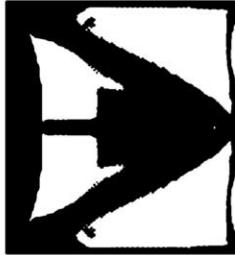 | 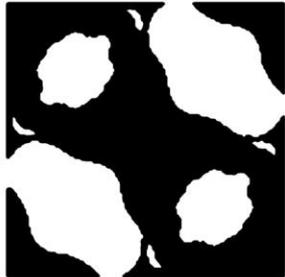 |
| 3.00E-07 | 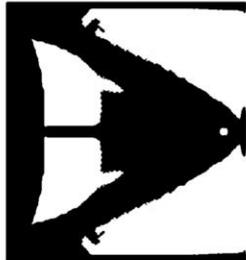 | 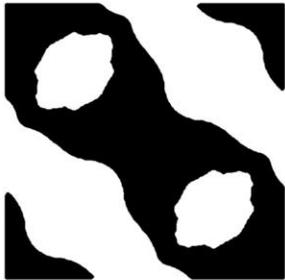 |

Table 14  Comparison of multiscale designs for laser-activated porous actuators incorporating a micro-scale with initial three-diamond-shaped voids, evaluated under different convection rates.

| Convcetion Rate | Macroscale design | Microscale design |
|---|---|---|
| 1.00E-07 | 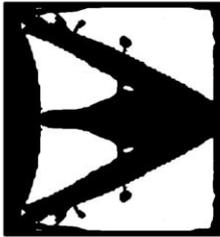 | 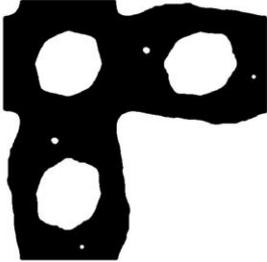 |
| 2.00E-07 | 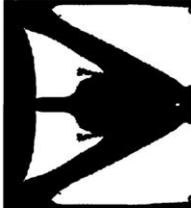 | 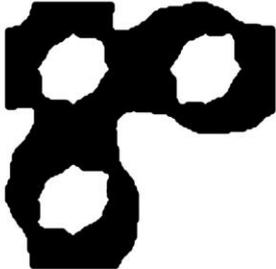 |
| 3.00E-07 | 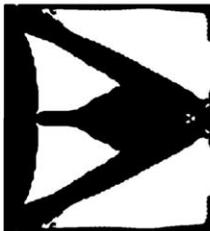 | 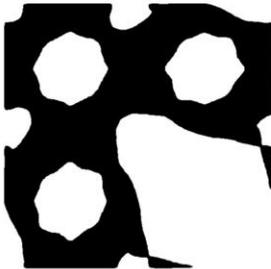 |

Table 15 Comparison of multiscale designs for laser-activated porous actuators incorporating a micro-scale with three or four initial voids, evaluated at different convection rates.

| Convcetion Rate | Macroscale design | Microscale design |
|---|---|---|
| 1.00E-07 | 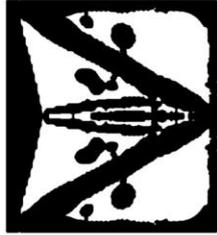 | 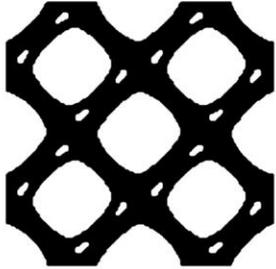 |
| 2.00E-07 | 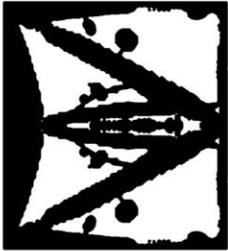 | 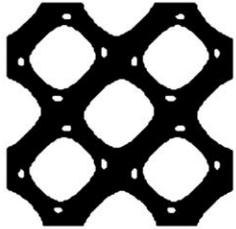 |
| 3.00E-07 | 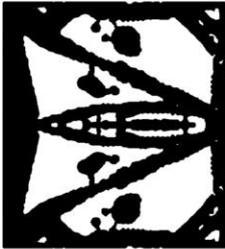 | 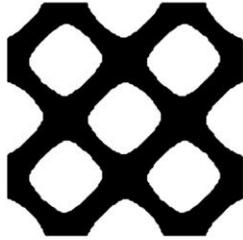 |

## 4. Conclusion

This study focuses on the application of multi-physics multiscale topology optimization for the development of a lightweight porous linear actuation mechanism activated by laser. The investigation meticulously explores the nuances of thermal dissipation mechanisms, incorporating the dynamics of conduction, convection, and radiation. The influence of microstructure systems on the topology optimization process is particularly intriguing. Various numerical cases are examined to discern the impact of microscopic-scale considerations on porous design, revealing a substantial 45% improvement in results compared to solid actuator designs. Furthermore, our investigation extends to the simultaneous optimization of multiscale porous displacement actuators, emphasizing a 75% weight reduction. Remarkably, the performance of the simultaneously optimized multiscale porous displacement actuator surpasses that of a comparable weight reduction in the case of a single scale. It demonstrates a significant improvement of 45.5% compared to the 50% weight reduction and a substantial improvement of 33.3% compared to the single-scale case with a 75% weight reduction. The increased freedom in the micro-scale design domain allows the simultaneous optimization algorithm to more efficiently distribute materials for the heat-activated porous displacement actuator. This influence is observed not only in the macro configuration but also in the overall layout, illustrating the effectiveness of the multiscale approach in optimizing the performance of laser-activated actuators. Our exploration delves into the sequential optimization of macro and micro-scale porous displacement actuators, contrasting with concurrent multiscale optimization. This comparative analysis aims to discern performance variations, especially considering the interconnected responses of both systems throughout the optimization process. Surprisingly, optimization accessible through separate macro and micro-scale optimizations yields inferior performance compared to simultaneous multiscale optimization. Expanding the scope of our investigation, we explore topology optimization under energy dissipation, focusing on multiple-rate thermal convection. The response to thermal convection is manifested by the development of extensions from the main design body, particularly evident at the macro scale of the single-scale optimized case. Extending this exploration to multiscale optimization reveals a parallel trend, with increased thermal convection leading to an extension of the main design body. This adaptive behavior emphasizes the role of convection in determining design resilience. Furthermore, our results reveal distinct patterns along the design boundaries, with pointed formations emerging and extending from the main body in response to intensified thermal convection. The macro-scale design, influenced by increased convective heat transfer, incorporates perpendicular links within its structure, causing corresponding adaptations in the microstructure arrangement. At the multiscale, in response, the shear modulus increases, while the Young's modulus decreases in the x and y directions. This adaptive strategy aims to enhance resilience and elasticity against heat-induced movements while simultaneously strengthening torsional resistance that could impede linear movement in the x direction; the objective of the optimization process. The solid thermal conductivity tensor within the microstructure reflects concerted efforts to reduce heat transfer to the newly formed extensions. Emphasizing the complex interplay between design adjustments and thermal convection dynamics, the expanded investigation aims to encompass energy dissipation through thermal radiation and convection, revealing a propensity to construct radiation dissipation surfaces directed towards space. Simultaneously, there is a strategic maintenance of a considerable surface to achieve balanced thermal dissipation through convection. The results highlight the intricate design considerations required to optimize thermal dissipation mechanisms, particularly in the context of constructing surfaces that effectively dissipate heat through radiation and convection processes.

**Competing interests**

The authors declare no competing financial interests or personal relationships that could have appeared to influence the work reported in this work.

**Acknowledgment**

Part of this work was supported by a Grant-in-Aid for Scientific Research awarded by the Japan Society for the Promotion of Science (JSPS), KAKEN of Grant Number JP21K03757.